\documentstyle[preprint,aps]{revtex}
\tightenlines              
\begin{document}

\title{Amplitude analysis of hadron decays}
\author{R. Delbourgo\cite{RD} and Dongsheng Liu\cite{DL}}
\address{Physics Department, University of Tasmania,
         GPO Box 252C, Australia 7001}
\date{\today}
\maketitle

\begin{abstract}
We provide succinct covariant amplitude decompositions of 2-body weak
hadronic decays, with which to compare data, including exclusive
rates, helicity amplitudes and polarizations. For weak decays,
the systematic dependence of these amplitudes on masses and quantum numbers
of participating particles are determined within a factor of about two by
the CKM angles and the Fermi constant so theoretical models need to be
much more accurate if they are to be convincing.
\end{abstract}
\pacs{???}

\section{Introduction}

Weak nonleptonic processes constitute a sizeable fraction of the decay
channels of recently discovered heavy charm and bottom mesons and baryons, 
not to mention strange hadrons. While a great deal is known about the 
semileptonic decays of those particles and much research \cite{semi} has been
devoted to determining the hadronic matrix elements of the weak currents,
we are rather less sure about the underlying picture \cite{nonold} when
{\em no} leptons emerge in the decay, because the strong interactions
influence significantly the effective weak interactions: the gluons tend to
muddy the current-current picture that governs semileptonic processes and it
is not even certain that a factorization model \cite{nonnew} of four-quark
operators prevails, especially for charmed states and heavy hadrons
\cite{4q}.

In order to gain some understanding of the dynamics governing nonleptonic 
decays it is helpful to express the decay amplitudes in some 
sort of standard kinematic form, with which to compare the theoretical
predictions. This paper is devoted to expressing all 2-body nonleptonic
amplitudes in the cleanest possible way, one which is neatly related to
helicity amplitude decompositions. We will carry out our analysis in a
covariant framework, since it is the easiest way to make contact with field
theoretic descriptions in terms of various operators; an added advantage is 
that one can cross over to other channels easily, via simple substitution
rules, and one need not worry about threshold factors since the latter
come in naturally after contraction over external wavefunctions, provided one
is careful not to introduce artificial kinematic singularities in the
covariant decompositions. In principle one could extend the analysis to
many-body non-resonant channels---abundant when the heavier hadrons decay---
but really one requires some detailed knowledge of the 2-body hadron dynamics
before one can make true progress on that wider front.

We shall consider the case of a particle/resonance of spin $j_1$, mass $m_1$ 
decaying into two lighter particles of spins and masses $j_1,j_2$ and 
$m_2,m_3$ respectively. The number of amplitudes $N$ required for 
describing the decay is given in the appendix and to each of them is
associated some set of coupling constants $g_i^{j_1j_2j_3}(m_1,m_2,m_3)$. 
If one plots the $g$ in a three-dimensional graph as a function of its
three mass arguments and because a physical process restricts us to the 
kinematic region $m_1>m_2+m_3$ (and its crossed versions, $m_2>m_3+m_1, 
m_3>m_1+m_2$), one will directly sample only a region of phase space---it lies 
outside the convex region bounded by the three planes joining pairs of lines 
$m_1=m_2, m_2=m_3, m_3=m_1$. We expect the couplings $g_i$ to depend quite
substantially on the masses, because they are connected through the quantum 
numbers of the constituent quarks, if nothing else.\footnote{Sometimes the
renormalization group can enlighten us about the variation with mass scale in a
scheme like the standard model.}  For large quantum number changes we
anticipate big variations in the $g_i$ through the character of the CKM
mixing matrix, but we expect the same thing to happen when higher radial
or angular quantum numbers are excited as a consequence of the basic quark
dynamics. There is quite of lot of data which lends iteslf to this kind of
systematic analysis; however, if one wants to get inside the convex mass
region, 2-body decays are useless: it becomes necessary to consider 3- or
higher-body processes where the strong dynamics is known to be dominated by
certain pole terms in various kinematical sections of the Dalitz plot and
extrapolate to those regions in order to pick out the residues and couplings.
It is worth mentioning that the $g_i$ entering the effective Lagrangians
cannot be real, because plain unitarity guarantees that their phase will be
connected to the scattering amplitude of the outgoing hadrons at the very
least. Nor does it follow that $g$ are necessarily real in the convex mass
region, for it is sometimes possible that strong intermediate states
(especially pions) arise in some decay channel below the exclusive
mass threshold. In what follows we shall be careful not to assume that the
$g$ are real.

We will focus attention on mesons with spins $j$ up to 1 and baryons with
spins up to 3/2, though it is quite easy to generalize the formulae below
to situations when the initial particle has higher spin. Throughout we
shall express the effective Lagrangian in the generic form
\[
 {\cal L}^{j_1j_2j_3}=g^{j_1j_2j_3}\phi_2^*(p_2)\phi_3^*(p_3)\phi_1(-p_1), 
\]
where we interpret $\phi$ as unity for spin 0, as the polarization vector
$\varepsilon^\mu$ for spin 1, as the standard Dirac spinor $u$ for spin 1/2
and as the Rarita-Schwinger spinor $u^\mu$ for spin 3/2; for antifermions
we replace incoming $u(p)$ by outgoing $v(p)$, in the standard way. It is
also understood that there may arise several independent tensors, with
associated coupling constants, which one may have to sum over. (See the
Appendix for notational details.) To reiterate, the couplings clearly 
depend on the participating particle masses which are intimately tied to their 
quantum numbers, including internal ones like flavour. The aim of the exercise 
is to extract the couplings from the available data and look for systematic
properties/deviations from symmetry predictions in order to pinpoint
afterwards the underlying forces capable of producing such effective
interactions.

The paper is divided into three parts. Section II deals with baryons: all
their covariant decompositions and connections with helicity amplitudes.
Section III contains the same tabulation for decaying mesons. Finally we look
at some typical magnitudes in section IV, not only to exhibit the 
mass-dependence/quantum number relation, but to make the point that once
phase-space and CKM angles are factored out, the couplings are all similar
to within a factor of 2; this is is ultimately tied to the 4-quark 
current-current picture and it signifies that experimental models should
aim for accuracy to better than 10\% if they are to claim our attention. 
The purpose of this paper is to set the groundwork for a full investigation 
of experimentally measured exclusive decay processes \cite{PD} and thereby 
unravel the full form of the {\em effective} Lagrangian. This
must then be converted into the form of multiquark operators, since this is
the preferred way of probing the quark dynamics. The conversion requires
specific knowledge of how the hadronic wavefunctions are built up from the
quark field operators at that level and some sort of tracing formula. Be that
as it may, the emergent multiquark operators encapsulate the combined effect
of weak {\em and} strong interactions.
        
\section{Baryon decay widths}

\subsection{Covariant Amplitudes}

We shall examine baryon decays where the outgoing baryon is identified with 
particle 2 and the meson with particle 3; in this paper we will focus on
partial decay widths, rather than more refined observables like angular
asymmetries and polarizations. We shall not assume anything about parity
conservation since we are primarily interested in weak decays. The effective
Lagrangian always takes the form,
${\cal L} = \phi^*(p_3)\bar{u}(p_2){\cal M}u(-p_1)$ and $\cal M$ will contain 
two sets of opposite parity couplings which we label $f$ and $g$; the latter 
multiplies a negative parity quantity like $\gamma_5$ or the Levi-Civita
symbol $\epsilon$. Remember that $-p_1$ is the physical incoming momentum,
which is somewhat unconventional.

\noindent
$\underline {1/2 \rightarrow 1/2 + 0}$

The effective Lagrangian equals
\[
 {\cal L}^{\frac{1}{2}\frac{1}{2} 0} = \bar{u}_2 (f^{\frac{1}{2}\frac{1}{2}0} 
                               + g^{\frac{1}{2}\frac{1}{2} 0} \gamma_5)u_{1}.
\]
The coupling constant $f^{\frac{1}{2}\frac{1}{2}0}$ of the scalar interaction 
is associated with an $s$-wave amplitude, while $g^{\frac{1}{2}\frac{1}{2}0}$ 
of the pseudoscalar interaction is tied to a $p$-wave one. 
In the resulting decay width,
\[
 \Gamma^{\frac{1}{2}\frac{1}{2}0} =\frac{\Delta}{8\pi m_1^3}  
 [(m_1m_2 - p_1\cdot p_2)|f^{\frac{1}{2}\frac{1}{2}0}|^2  -
 (m_1m_2 + p_1\cdot p_2)|g^{\frac{1}{2}\frac{1}{2}0}|^2],
\]
\[
 {\rm where} \quad \Delta^2\equiv [m_1^2-(m_2+m_3)^2][m_1^2-(m_2-m_3)^2],
\]
note that one can obtain the contribution of $g^{\frac{1}{2}\frac{1}{2}0}$ 
from that of $f^{\frac{1}{2}\frac{1}{2}0}$ by substituting $-m_1m_2$ for
$m_1m_2$, with no interference between them.
This is a recurrent feature. 

Among a plethora of such decays are typically, $\Lambda \rightarrow N\pi$
and $\Lambda_c \rightarrow \Sigma\pi$.

\noindent
$\underline {1/2 \rightarrow 1/2 + 1}$

We write the interaction Lagrangian in the Sachs form,
\[
 {\cal L}^{\frac{1}{2}\frac{1}{2}1} = \varepsilon_3^{\nu *}\bar u_2 \left[ 
 \frac{1}{2}(p_2-p_1)_\nu(f_E^{\frac{1}{2}\frac{1}{2}1} +
                            g_E^{\frac{1}{2}\frac{1}{2}1}\gamma_5) +
         \epsilon_{\nu\alpha\beta\rho}p_1^\alpha p_2^\beta\gamma^\rho 
      (f_M^{\frac{1}{2}\frac{1}{2}1}\gamma_5+g_M^{\frac{1}{2}\frac{1}{2}1}) 
 \right] u_{1}. 
\]
The corresponding decay width assumes a pleasing diagonal form\footnote{If 
one invokes the standard vector $\gamma_\mu$ and axial vector 
$\gamma_\mu\gamma_5$ couplings plus magnetic moment $\sigma_{\mu\nu}p_3^\nu$
counterpart terms, instead of the Sachs kinematic covariants, interference
terms abound.},
\[
 \Gamma^{\frac{1}{2}\frac{1}{2}1}\!=\!\frac{\Delta^3}{32\pi m_1^3} \left[
 (m_1m_2\!-\!p_1\!\cdot\!p_2)
 \left(\frac{|f_E^{\frac{1}{2}\frac{1}{2}1}|^2}{m_3^2}\!+
 2|f_M^{\frac{1}{2}\frac{1}{2}1}|^2 \right)\! - \!
 (m_1m_2\!+\!p_1\!\cdot\!p_2) 
 \left(\frac{|g_E^{\frac{1}{2}\frac{1}{2}1}|^2}{m_3^2}\!+
 2|g_M^{\frac{1}{2}\frac{1}{2}1}|^2\right)\! \right] 
\]
When particle 3 is a photon, and the baryon masses $m_1$ and $m_2$ are 
unequal as is necessary for the decay to occur, gauge invariance requires 
$f_E^{\frac{1}{2}\frac{1}{2}1}=g_E^{\frac{1}{2}\frac{1}{2}1}\rightarrow 0$. 
As a result the radiative width simplifies to 
\[
 \Gamma^{\frac{1}{2}\frac{1}{2}1} \rightarrow 
 \frac{(m_1^2-m_2^2)^3}{16\pi m_1^3} \left[
 (m_1 + m_2)^2|f_M^{\frac{1}{2}\frac{1}{2}1}|^2 +
 (m_1 - m_2)^2|g_M^{\frac{1}{2}\frac{1}{2}1}|^2 \right], 
\]
with $f_M^{\frac{1}{2}\frac{1}{2}1}$ or $g_M^{\frac{1}{2}\frac{1}{2}1}$ 
coupling disappearing, depending on the relative parity between the baryons.

Examples of such processes are $\Sigma^+\rightarrow p\gamma, \Lambda_b
\rightarrow \psi\Lambda, \Lambda_c\rightarrow p \phi$.

\noindent
$\underline {3/2 \rightarrow 1/2 + 0}$

Moving on to excited baryon decays, this first case involves two couplings
corresponding to $d$ and $p$ wave amplitudes. The effective interaction can 
be written uniquely as  
\[
 {\cal L}^{\frac{3}{2}\frac{1}{2}0} = 
 \frac{1}{2}(p_3-p_2)_\lambda\bar{u}_2(f^{\frac{3}{2}\frac{1}{2}0} 
      + g^{\frac{3}{2}\frac{1}{2}0}\gamma_5) u_{1}^\lambda,
\]
and it is straightforward to calculate the decay width, 
\[
 \Gamma^{\frac{3}{2}\frac{1}{2}0}=\frac{\Delta^3}{96\pi m_1^5} \left[
 (m_1m_2-p_1\cdot p_2)|f^{\frac{3}{2}\frac{1}{2}0}|^2  -
 (m_1m_2+p_1\cdot p_2)|g^{\frac{3}{2}\frac{1}{2}0}|^2 \right]. 
\]

The decays $\Omega \rightarrow \Lambda K$ and $\Xi^* \rightarrow\Lambda\pi$
illustrate this case.

\noindent
$\underline {1/2 \rightarrow 3/2 + 0}$

This is basically the same as the previous case, since it corresponds to 
the interchange $p_1 \leftrightarrow p_2$, apart from the initial spin
averaging factor. (Note that the triangular Kallen function $\Delta$ 
is symmetric under mass exchange.) 

An example is $N(1440) \rightarrow \Delta\pi.$

\noindent
$\underline {3/2 \rightarrow 1/2 + 1}$

When the final meson is vectorial we come across six independent amplitudes, 
for which we adopt the following Lagrangian ,
\[
\begin{array}{rl}
 {\cal L}^{\frac{3}{2}\frac{1}{2}1} = \varepsilon^{\nu *}_3\bar{u}_2 & 
 [ (p_1\cdot p_3 g_{\lambda\nu}-p_{3\lambda}p_{1\nu})
 (f_T^{\frac{3}{2}\frac{1}{2}1}+g_T^{\frac{3}{2}\frac{1}{2}1}\gamma_5)
  + p_{3\lambda} p_{1\nu} 
  (f_L^{\frac{3}{2}\frac{1}{2}1}+g_L^{\frac{3}{2}\frac{1}{2}1}\gamma_5) \cr
 & +\frac{1}{2}(p_3-p_2)_\lambda
 \epsilon_{\nu\alpha\beta\rho}p_2^\alpha p_3^\beta\gamma^\rho
 (f_M^{\frac{3}{2}\frac{1}{2}1}\gamma_5+g_M^{\frac{3}{2}\frac{1}{2}1}) ] 
 u_{1}^\lambda .
\end{array}
\]
The decay width is quasidiagonal:
\[
\begin{array}{ll}
\displaystyle \Gamma^{\frac{3}{2}\frac{1}{2}1} = \frac{\Delta(m_1m_2 -
 p_1\cdot p_2)} {96\pi m_1^3} 
 & \left[ 2(\Delta^2+6m_1^2m_3^2)|f_T^{\frac{3}{2}\frac{1}{2}1}|^2 
 +\frac{\Delta^4}{4m_1^2m_3^2}|f_L^{\frac{3}{2}\frac{1}{2}1}|^2 
 +\frac{\Delta^4}{2m_1^2} |f_M^{\frac{3}{2}\frac{1}{2}1}|^2 \right.\cr 
 & \left. +2\Delta^2{\rm Re}
 (f_T^{\frac{3}{2}\frac{1}{2}1}f_L^{\frac{3}{2}\frac{1}{2}1*})
 -2\frac{\Delta^2}{m_1} p_1\!\cdot\!p_3 
 {\rm Re}(if_T^{\frac{3}{2}\frac{1}{2}1^*}f_M^{\frac{3}{2}\frac{1}{2}1}) 
 \right] \cr
 \qquad + \,\,m_1\leftrightarrow -m_1, f_i\leftrightarrow g_i & .
\end{array}
\]
This result simplifies when a final state photon is produced because
gauge invariance leads to $f_L^{\frac{3}{2}\frac{1}{2}1} =
g_L^{\frac{3}{2}\frac{1}{2}1}\rightarrow 0$. We end up with the expression,
\[
\begin{array}{rl}
 \displaystyle \Gamma^{\frac{3}{2}\frac{1}{2}1}\!\rightarrow\!
 \frac{(m_1^2-m_2^2)^3}{384\pi m_1^3} & \displaystyle
 \!\left\{\!(m_1\!+\!m_2)^2\!\left[|f_T^{\frac{3}{2}\frac{1}{2}1}|^2\! +
 \frac{(m_1^2\!-\!m_2^2)^2}{4m_1^2}|f_M^{\frac{3}{2}\frac{1}{2}1}|^2\! -
 \frac{m_1^2\!-\!m_2^2}{2m_1} {\rm Re}(if_M^{\frac{3}{2}\frac{1}{2}1}
 f_T^{\frac{3}{2}\frac{1}{2}1*})\!\right] \right. \cr 
&\displaystyle\left.+(m_1\!-\!m_2)^2\!\left[|g_T^{\frac{3}{2}\frac{1}{2}1}|^2\!
 +\frac{(m_1^2\!-\!m_2^2)^2}{4m_1^2}|g_M^{\frac{3}{2}\frac{1}{2}1}|^2\! -
 \frac{m_1^2\!-\!m_2^2}{2m_1}{\rm Re}(ig_M^{\frac{3}{2}\frac{1}{2}1}
 g_T^{\frac{3}{2}\frac{1}{2}1*})\!\right]\!\right\}
\end{array}
\]
Examples of such decays are $\Delta \rightarrow N\gamma,
\Xi^*\rightarrow\Xi\gamma.$.

\noindent
$\underline {1/2 \rightarrow 3/2 + 1}$

This is basically the same as the previous case with permutation of momenta
$p_1$ and $p_3$, but examples of this process from particle physics are hard
to find. One which has been seen is $\Lambda_c \rightarrow\Delta\bar{K}^*$.

\noindent
$\underline {3/2 \rightarrow 3/2 + 0}$

The last two cases we consider are transitions from excited to excited
baryons. When the final meson has spin zero, the Lagrangian takes the form,
\[
 {\cal L}^{\frac{3}{2}\frac{3}{2}0}=\bar{u}_2^\mu  
 [(f_T^{\frac{3}{2}\frac{3}{2}0}+g_T^{\frac{3}{2}\frac{3}{2}0}\gamma_5)
 (p_1\cdot p_2 g_{\mu\lambda}-p_{1\mu}p_{2\lambda}) +
 (f_L^{\frac{3}{2}\frac{3}{2}0}+g_L^{\frac{3}{2}\frac{3}{2}0}\gamma_5)
 p_{1\mu}p_{2\lambda}]u_{1}^\lambda, 
\]
and the width is determined to be 
\[
\begin{array}{ll}
 \displaystyle \Gamma^{\frac{3}{2}\frac{3}{2}0}=
 \frac{\Delta}{144\pi m_1^3}(m_1m_2-\!p_1\!\cdot\! p_2\!)
 &\left\{ \displaystyle [9(p_1\!\cdot\!p_2)^2 +
 (2m_1m_2-p_1\!\cdot\!p_2)^2]|f_T^{\frac{3}{2}\frac{3}{2}0}|^2 +\right.\cr
 & \displaystyle\left. \frac{\Delta^4}{4m_1^2m_2^2}
 |f_L^{\frac{3}{2}\frac{3}{2}0}|^2+\frac{\Delta^2}{m_1m_2}
 (2m_1m_2-p_1\!\cdot\!p_2\!){\rm Re}
 (f_T^{\frac{3}{2}\frac{3}{2}0}f_L^{\frac{3}{2}\frac{3}{2}0*})
 \right\}\cr
 & + m_1m_2\leftrightarrow -m_1m_2,\qquad f_i\leftrightarrow g_i.
\end{array}
\]

Examples of these processes are rare. One is $N(1520) \rightarrow \Delta \pi$.

\noindent
$\underline {3/2 \rightarrow 3/2 + 1}$

This vector channel brings in ten independent amplitudes. We write the 
effective interaction Lagrangian as 
\[
\begin{array}{ll}
 {\cal L}^{\frac{3}{2}\frac{3}{2}1}=\varepsilon^{\nu *}_3\bar{u}_2^\mu & 
 \left\{\frac{1}{2}(p_3-p_2)_\lambda(f_{1T}^{\frac{3}{2}\frac{3}{2}1}+
 g_{1T}^{\frac{3}{2}\frac{3}{2}1} \gamma_5)
 (p_2\cdot p_3g_{\mu\nu}-p_{2\nu}p_{3\mu}) 
    \right. \cr
 &\left. +\frac{1}{2}(p_1-p_3)_\mu
 (f_{2T}^{\frac{3}{2}\frac{3}{2}1}+g_{2T}^{\frac{3}{2}\frac{3}{2}1}\gamma_5) 
 (p_1\cdot p_3g_{\nu\lambda}-p_{1\nu}p_{3\lambda})\right. \cr
 &\left.+\frac{1}{2}(p_2-p_1)_\nu
   (f_{3T}^{\frac{3}{2}\frac{3}{2}1}+ g_{3T}^{\frac{3}{2}\frac{3}{2}1}\gamma_5) 
    (p_1\cdot p_2 g_{\mu\lambda}-p_{1\mu}p_{2\lambda})
\right. \cr
 &\left.+\frac{1}{8}(p_3-p_2)_\lambda (p_1-p_3)_\mu (p_2-p_1)_\nu  
 (f_L^{\frac{3}{2}\frac{3}{2}1}+g_L^{\frac{3}{2}\frac{3}{2}1}\gamma_5)\right. \cr
 &\left. +g_{\mu\lambda}
         \epsilon_{\nu\alpha\beta\rho} p_1^\alpha p_2^\beta \gamma^\rho
        (f^{\frac{3}{2}\frac{3}{2}1}_M\gamma_5+g_M^{\frac{3}{2}\frac{3}{2}1})
 \right\} u_{1}^\lambda .
 \end{array}
\]
It is straightforward, though tedious, to work out the decay width which
here reads 
\[
\begin{array}{ll}\displaystyle 
 \Gamma^{\frac{3}{2}\frac{3}{2}1}= 
 \frac{\Delta^3(m_1m_2-p_1\cdot p_2)}{144\pi m_1^3} \!\!&\!\!
 \left\{\displaystyle 
 \frac{1}{2m_1^2}(\Delta^2+6m_2^2m_3^2)|f_{1T}^{\frac{3}{2}\frac{3}{2}1}|^2
+\frac{1}{2m_2^2}(\Delta^2+6m_1^2m_3^2)|f_{2T}^{\frac{3}{2}\frac{3}{2}1}|^2
\right.\cr & \displaystyle 
+\frac{1}{4m_3^2}[9(p_1\cdot p_2)^2+(2m_1m_2-p_1\cdot p_2)^2] 
 |f_{3T}^{\frac{3}{2}\frac{3}{2}1}|^2
\cr & \displaystyle 
+\frac{\Delta^4}{16m_1^2m_2^2m_3^2} |f_L^{\frac{3}{2}\frac{3}{2}1}|^2
+\frac{1}{2}\left(10+\frac{\Delta^2}{m_1^2m_2^2}\right)
                 |f_M^{\frac{3}{2}\frac{3}{2}1}|^2 -
\cr & \displaystyle 
 m_1m_2(2m_1m_2\!-\!p_1\!\cdot\!p_2) 
 {\rm Re}\!\left[\left(\frac{f_{1T}^{\frac{3}{2}\frac{3}{2}1}}{m_1^2}\!
                  \!+\frac{f_{2T}^{\frac{3}{2}\frac{3}{2}1}}{m_2^2}\!
                  \!-\!\frac{\Delta^2}{4m_1^2m_2^2}
                     \frac{f_L^{\frac{3}{2}\frac{3}{2}1}}{m_3^2}\right)
                         \!\!f_{3T}^{\frac{3}{2}\frac{3}{2}1 *}\right]
\cr & \displaystyle 
-\left(3-2\frac{p_1\cdot p_2 }{m_1m_2}\right) 
 {\rm Re}\left[if_M^{\frac{3}{2}\frac{3}{2}1}
 \left(\frac{p_2\cdot p_3}{m_1}f_{1T}^{\frac{3}{2}\frac{3}{2}1}
+\frac{p_1\cdot p_3}{m_2}f_{2T}^{\frac{3}{2}\frac{3}{2}1}\right)^*\right]
\cr & \displaystyle
-\frac{\Delta^2-4m_3^2(p_1\cdot p_2+2m_1m_2)}{4m_1m_2}
 {\rm Re}(f_{1T}^{\frac{3}{2}\frac{3}{2}1}f_{2T}^{\frac{3}{2}\frac{3}{2}1*})
\cr & \displaystyle \left. 
-\frac{\Delta^2}{2}{\rm Re}\left[\left(
 \frac{f_{1T}^{\frac{3}{2}\frac{3}{2}1}}{m_1^2}
+\frac{f_{2T}^{\frac{3}{2}\frac{3}{2}1}}{m_2^2}\!\right)
          f_L^{\frac{3}{2}\frac{3}{2}1 *}\right] 
\right\} \cr & 
+ m_1m_2\leftrightarrow -m_1m_2,\quad f_i\leftrightarrow g_i. 
\end{array}
\]
In the massless vector limit, gauge invariance gives rise to conditions
$f^{\frac{3}{2}\frac{3}{2}1}=f_L^{\frac{3}{2}\frac{3}{2}1}=
g^{\frac{3}{2}\frac{3}{2}1}=g_L^{\frac{3}{2}\frac{3}{2}1}\rightarrow 0$,
so the decay width reduces to
\[
\begin{array}{ll}\displaystyle
 \Gamma^{\frac{3}{2}\frac{3}{2}1} \rightarrow
 \frac{(m_1^2-m_2^2)^3(m_1+m_2)^2}{576\pi m_1^3} & \displaystyle
 \left\{(m_1^2-m_2^2)^2\left(\frac{|f_{1T}^{\frac{3}{2}\frac{3}{2}1}|^2}{m_1^2}
                            +\frac{|f_{2T}^{\frac{3}{2}\frac{3}{2}1}|^2}{m_2^2}
                       \right)
 \right.\cr
 & \!\displaystyle
+\left(8+\frac{m_1^2}{m_2^2}+\frac{m_2^2}{m_1^2}\right)
 |f_M^{\frac{3}{2}\frac{3}{2}1}|^2
-\frac{(m_1^2-m_2^2)^2}{2m_1m_2}
 {\rm Re}(f_{1T}^{\frac{3}{2}\frac{3}{2}1}f_{2T}^{\frac{3}{2}\frac{3}{2}1*})
\cr & \!\displaystyle\left.
\!-\!(m_1^2-m_2^2)\!\left(\!3\!+\!\frac{m_1}{m_2}+\!\frac{m_2}{m_1}\right) 
 \!{\rm Re}\!\left[\! if_M^{\frac{3}{2}\frac{3}{2}1}\!
 \left(\frac{f_{1T}^{\frac{3}{2}\frac{3}{2}1}}{m_1}\!
 -\!\frac{f_{2T}^{\frac{3}{2}\frac{3}{2}1}}{m_2}\right)^* \right]\!
\right\}\cr
 & +m_1m_2\leftrightarrow -m_1m_2,\quad f_i\leftrightarrow g_i.
\end{array}
\]

We have been unable to spot any examples of such decays from Particle
Data Tables.

\subsection{Helicity amplitudes}

The appendix contains the definition of the reduced helicity amplitudes
$M_{\lambda_2,\lambda_3}$. They can be calculated in terms of the covariantly 
defined coupling constants $f^{j_1j_2j_3}$ and $g^{j_1j_2j_3}$ above by 
proceeding to the centre of mass frame where,
\[
 -p_1 = (m_1;0,0,0),\quad p_2=(E_2;q\sin\theta,0,q\cos\theta), \quad
  p_3 = (E_3;-q\sin\theta,0,-q\cos\theta),
\]
and
\[ 
2m_1E_2=m_1^2+m_2^2-m_3^2,\quad 2m_1E_3=m_1^2+m_3^2-m_2^2,\quad
2m_1q = \Delta(m_1^2,m_2^2,m_3^2).
\]
One then carries out the contraction over wavefunctions for particular
helicities, as needed, and compares against the expected result, viz.
$d^{j_1}_{\lambda_1\,\lambda_2-\lambda_3}(\theta)M_{\lambda_2,\lambda_3}.$ 
For that purpose one requires the polarization vectors
\[
\varepsilon^{(\pm 1)}(-p_1)=(0;\mp 1,-i,0)/\sqrt{2},\quad
\varepsilon^{(0)}(-p_1)=(0;0,0,1),
\]
\[
\varepsilon^{(\pm 1)*}(p_2)=\varepsilon^{(\mp 1)*}(p_3)
 =(0;\mp\cos\theta,i,\pm\sin\theta)/\sqrt{2},
\]
\[
\varepsilon^{(0)}(p_2)=(q;E_2\sin\theta,0,E_2\cos\theta)/m_2, \quad
\varepsilon^{(0)}(p_3)=(q;-E_3\sin\theta,0,-E_3\cos\theta)/m_3.
\]
(Obviously we disregard the last two zero helicity vectors when photons
are involved.)

In addition we shall need the specific forms for the particle and
antiparticle wavefunctions, $u$ and $v$ respectively, in the centre of mass
frame, when considering the decays of mesons into two fermions. In our
notation, the transposes (\verb+~+) read:
\[
\tilde{u}^{( 1/2)}(-p_1)=\sqrt{2m_1}(1,0,0,0); \quad
\tilde{u}^{(-1/2)}(-p_1)=\sqrt{2m_1}(0,1,0,0);
\]
\[
\tilde{u}^{(1/2)}(p_2)=(\sqrt{E_2+m_2}\cos\frac{\theta}{2},
                          \sqrt{E_2+m_2}\sin\frac{\theta}{2},
                          \sqrt{E_2-m_2}\cos\frac{\theta}{2},
                          \sqrt{E_2-m_2}\sin\frac{\theta}{2});
\]
\[
\tilde{u}^{(-1/2)}(p_2)=(-\sqrt{E_2+m_2}\sin\frac{\theta}{2},
                          \sqrt{E_2+m_2}\cos\frac{\theta}{2},
                          \sqrt{E_2-m_2}\sin\frac{\theta}{2},
                          -\sqrt{E_2-m_2}\cos\frac{\theta}{2});
\]
\[
\tilde{v}^{(1/2)}(p_3)=(-\sqrt{E_3-m_3}\cos\frac{\theta}{2},
                           -\sqrt{E_3-m_3}\sin\frac{\theta}{2},
                           \sqrt{E_3+m_3}\cos\frac{\theta}{2},
                           \sqrt{E_3+m_3}\sin\frac{\theta}{2});
\]
\[
\tilde{v}^{(-1/2)}(p_3)=(\sqrt{E_3-m_3}\sin\frac{\theta}{2},
                          -\sqrt{E_3-m_3}\cos\frac{\theta}{2},
                          \sqrt{E_3+m_3}\sin\frac{\theta}{2},
                          -\sqrt{E_3+m_3}\cos\frac{\theta}{2}).
\]
We now have all the ammunition needed to extract the helicity-coupling 
relations, which are written below in specific parity combinations for 
convenience.

\noindent
$\underline {1/2 \rightarrow 1/2 + 0}$

The symmetric combination of helicity amplitudes under $\lambda_2,
\lambda_3\leftrightarrow -\lambda_2, -\lambda_3$ (which is connected with
particular parity) takes the form 
\[
 \displaystyle \frac{1}{2}(M_{1/2,0}+M_{-1/2,0}) =
 \sqrt{2(m_1m_2-p_1\cdot p_2)}\;f^{\frac{1}{2}\frac{1}{2}0}, 
\]
while the antisymmetric combination (with opposite parity) reads 
\[
 \displaystyle \frac{1}{2}(M_{1/2,0}-M_{-1/2,0}) =
 -\sqrt{-2(m_1m_2+p_1\cdot p_2)}\;g^{\frac{1}{2}\frac{1}{2}0}.
\]
Remember that $m_1m_2+p_1\cdot p_2$ is negative in our convention: 
all momenta outgoing.

\noindent
$\underline {1/2 \rightarrow 1/2 + 1}$
\[\begin{array}{l}\displaystyle 
 \frac{1}{2}(M_{1/2,0}+M_{-1/2,0}) 
 =\sqrt{2(m_1m_2-p_1\cdot p_2)}\;\frac{\Delta}{2m_3} 
 f_E^{\frac{1}{2}\frac{1}{2}1}, 
\cr\cr\displaystyle   \frac{1}{2}(M_{1/2,1}+M_{-1/2,-1}) 
=-\sqrt{(m_1m_2-p_1\cdot p_2)}\;\Delta f_M^{\frac{1}{2}\frac{1}{2}1}, 
\cr\cr\displaystyle \frac{1}{2}(M_{1/2,0}-M_{-1/2,0}) 
=i\sqrt{-2(m_1m_2+p_1\cdot p_2)}\;\frac{\Delta}{2m_3} 
  g_E^{\frac{1}{2}\frac{1}{2}1}, 
\cr\cr\displaystyle \frac{1}{2}(M_{1/2,1}-M_{-1/2,-1}) 
=i\sqrt{-(m_1m_2+p_1\cdot p_2)}\;\Delta g_M^{\frac{1}{2}\frac{1}{2}1}. 
\end{array}\]

\noindent
$\underline {3/2 \rightarrow 1/2 + 0}$
\[\displaystyle 
 \frac{1}{2}(M_{1/2,0}+M_{-1/2,0}) =
\sqrt{\frac{1}{3}(m_1m_2-p_1\cdot p_2)}\;\frac{\Delta}{m_1}
f^{\frac{3}{2}\frac{1}{2}0}, 
\]
\[
 \displaystyle  \frac{1}{2}(M_{1/2,0}-M_{-1/2,0}) =i
\sqrt{-\frac{1}{3}(m_1m_2+p_1\cdot p_2)}\;\frac{\Delta}{m_1}
g^{\frac{3}{2}\frac{1}{2}0}.
\]
\noindent
$\underline {1/2 \rightarrow 3/2 + 0}$
\[\displaystyle 
 \frac{1}{2}(M_{1/2,0}+M_{-1/2,0}) =-
\sqrt{\frac{1}{3}(m_1m_2-p_1\cdot p_2)}\;\frac{\Delta}{m_2}
 f^{\frac{1}{2}\frac{3}{2}0}, 
\]
\[
 \displaystyle \frac{1}{2}(M_{1/2,0}-M_{-1/2,0}) =-i
\sqrt{-\frac{1}{3}(m_1m_2+p_1\cdot p_2)}\;\frac{\Delta}{m_2}
  g^{\frac{1}{2}\frac{3}{2}0}.
\]

\noindent
$\underline {3/2 \rightarrow 1/2 + 1}$

The symmetric combinations of helicity amplitudes are
\[\begin{array}{l}\displaystyle 
 \frac{1}{2}(M_{1/2,-1}+M_{-1/2,1}) = 
 -\sqrt{2(m_1m_2-p_1\cdot p_2)}\;p_1\cdot p_3f_T^{\frac{3}{2}\frac{1}{2}1}, 
 \cr\cr\displaystyle 
 \frac{1}{2}(M_{1/2,1}+M_{-1/2,-1}) = 
 -\sqrt{\frac{2}{3}(m_1m_2-p_1\cdot p_2)}\;
  \left(p_1\cdot p_3f_T^{\frac{3}{2}\frac{1}{2}1} 
 +\frac{\Delta^2}{2m_1}f_M^{\frac{3}{2}\frac{1}{2}1}\right), 
\cr\cr\displaystyle 
\frac{1}{2}(M_{1/2,0}+M_{-1/2,0}) = 
 -\sqrt{\frac{4}{3}(m_1m_2-p_1\cdot p_2)}\;
  \left(m_1m_3f_T^{\frac{3}{2}\frac{1}{2}1}
 +\frac{\Delta^2}{4m_1m_3}f_L^{\frac{3}{2}\frac{1}{2}1}\right), 
\end{array}\]
while the asymmetric ones are 
\[\begin{array}{l}\displaystyle
 \frac{1}{2}(M_{1/2,-1}-M_{-1/2,1}) = 
 -i\sqrt{-2(m_1m_2+p_1\cdot p_2)}\;p_1\cdot p_3g_T^{\frac{3}{2}\frac{1}{2}1},
\cr\cr\displaystyle 
 \frac{1}{2}(M_{1/2,1}-M_{-1/2,-1}) = 
 -i\sqrt{-\frac{2}{3}(m_1m_2+p_1\cdot p_2)}\;  
  (p_1\cdot p_3 g_T^{\frac{3}{2}\frac{1}{2}1} 
 +\frac{\Delta^2}{2m_1}g_M^{\frac{3}{2}\frac{1}{2}1}), 
\cr\cr\displaystyle 
 \frac{1}{2}(M_{1/2,0}-M_{-1/2,0}) = 
-i\sqrt{-\frac{4}{3}(m_1m_2+p_1\cdot p_2)}\; 
 \left(m_1m_3 g_T^{\frac{3}{2}\frac{1}{2}1}
 +\frac{\Delta^2}{4m_1m_3}g_L^{\frac{3}{2}\frac{1}{2}1}\right).
\end{array}\]

\noindent
$\underline {1/2 \rightarrow 3/2 + 1}$

The various well-defined parity combinations are:

\[\begin{array}{l}\displaystyle 
\frac{1}{2}(M_{3/2,1}+M_{-3/2,-1}) =
 \sqrt{2(m_1m_2-p_1\cdot p_2)}\; p_2\cdot p_3 f_T^{\frac{1}{2}\frac{3}{2}1}, 
\cr\cr\displaystyle 
\frac{1}{2}(M_{1/2,1}+M_{-1/2,-1}) =
 \sqrt{\frac{2}{3}(m_1m_2-p_1\cdot p_2)}\;
 \left(p_2\cdot p_3f_T^{\frac{1}{2}\frac{3}{2}1} 
 -\frac{\Delta^2}{2m_2}f_M^{\frac{1}{2}\frac{3}{2}1}\right),
\cr\cr\displaystyle 
\frac{1}{2}(M_{1/2,0}+M_{-1/2,0}) =
 \sqrt{\frac{4}{3}(m_1m_2-p_1\cdot p_2)}\;
 \left(m_2 m_3f_T^{\frac{1}{2}\frac{3}{2}1} 
 +\frac{\Delta^2}{4m_2m_3}f_L^{\frac{1}{2}\frac{3}{2}1}\right),
\end{array}\]

\[\begin{array}{l}\displaystyle 
\frac{1}{2}(M_{3/2,1}-M_{-3/2,-1}) =
 i\sqrt{-2(m_1m_2+p_1\cdot p_2)}\; p_2\cdot p_3 g_T^{\frac{1}{2}\frac{3}{2}1},
\cr\cr\displaystyle 
\frac{1}{2}(M_{1/2,1}-M_{-1/2,-1}) =
 -i\sqrt{-\frac{2}{3}(m_1m_2+p_1\cdot p_2)}\;
 \left(p_2\cdot p_3 g_T^{\frac{1}{2}\frac{3}{2}1} 
 +\frac{\Delta^2}{2m_2}g_M^{\frac{1}{2}\frac{3}{2}1}\right), 
\cr\cr\displaystyle 
\frac{1}{2}(M_{1/2,0}-M_{-1/2,0}) =
 i\sqrt{-\frac{4}{3}(m_1m_2+p_1\cdot p_2)}\;
  \left( m_2 m_3 g_T^{\frac{1}{2}\frac{3}{2}1} 
 +\frac{\Delta^2}{4m_2m_3}g_L^{\frac{1}{2}\frac{3}{2}1}\right).
\end{array}\]

\noindent
$\underline {3/2 \rightarrow 3/2 + 0}$
\[\begin{array}{l}\displaystyle
\frac{1}{2}(M_{3/2,0}+M_{-3/2,0})=-
 \sqrt{2(m_1m_2-p_1\cdot p_2)}\;  p_1\cdot p_2 f_T^{\frac{3}{2}\frac{3}{2}0}, 
\cr\cr\displaystyle 
\frac{1}{2}
(M_{1/2,0}+M_{-1/2,0})=-\sqrt{\frac{2}{9}(m_1m_2-p_1\cdot p_2)}
\left( (p_1\cdot p_2-2m_1m_2)f_T^{\frac{3}{2}\frac{3}{2}0}
-\frac{\Delta^2}{2m_1m_2}f_L^{\frac{3}{2}\frac{3}{2}0}\right),
\end{array}\]
\[\begin{array}{l}\displaystyle
\frac{1}{2}(M_{3/2,0}-M_{-3/2,0})=-i 
 \sqrt{-2(m_1m_2+p_1\cdot p_2)}\; p_1\cdot p_2 g_T^{\frac{3}{2}\frac{3}{2}0}, 
\cr\cr\displaystyle 
\frac{1}{2}
(M_{1/2,0}-M_{-1/2,0})=i\sqrt{-\frac{2}{9}(m_1m_2+p_1\cdot p_2)}
 \left((p_1\cdot p_2+2m_1m_2)g_T^{\frac{3}{2}\frac{3}{2}0}
+\frac{\Delta^2}{2m_1m_2}g_L^{\frac{3}{2}\frac{3}{2}0}\right).
\end{array}\]
\noindent
$\underline {3/2 \rightarrow 3/2 + 1}$

\[
\begin{array}{l}\displaystyle 
 \frac{1}{2}(M_{3/2,1}+M_{-3/2,-1}) =\sqrt{\frac{1}{3}(m_1m_2-p_1\cdot p_2)}\;
 \Delta\left(\frac{p_2\cdot p_3}{m_1} f_{1T}^{\frac{3}{2}\frac{3}{2}1}
            +f_M^{\frac{3}{2}\frac{3}{2}1}\right),
\cr\cr\displaystyle 
 \frac{1}{2}(M_{1/2,-1}+M_{-1/2,1})=\sqrt{\frac{1}{3}(m_1m_2-p_1\cdot p_2)}\;
 \Delta\left(\frac{p_1\cdot p_3}{m_2} f_{2T}^{\frac{3}{2}\frac{3}{2}1}
            +f_M^{\frac{3}{2}\frac{3}{2}1}\right),
\cr\cr\displaystyle 
 \frac{1}{2}(M_{1/2,1}+M_{-1/2,-1})=\sqrt{\frac{1}{9}(m_1m_2-p_1\cdot p_2)}\;
 \Delta\left(\frac{p_2\cdot p_3}{m_1}f_{1T}^{\frac{3}{2}\frac{3}{2}1}
            +\frac{p_1\cdot p_3}{m_2}f_{2T}^{\frac{3}{2}\frac{3}{2}1}
            -2\frac{p_1\cdot p_2}{m_1m_2}f_M^{\frac{3}{2}\frac{3}{2}1}\right), 
\cr\cr\displaystyle 
\frac{1}{2}(M_{3/2,0}+M_{-3/2,0})=
 -\sqrt{2(m_1m_2-p_1\cdot p_2)}\; \frac{\Delta}{2m_3} p_1\cdot p_2
  f_{3T}^{\frac{3}{2}\frac{3}{2}1}, 
\cr\cr\displaystyle  
\begin{array}{rl}\displaystyle  
 \frac{1}{2}(M_{1/2,0}+M_{-1/2,0})& \displaystyle  
=\sqrt{\frac{2}{9}(m_1m_2-p_1\!\cdot\!p_2)}\;  m_1m_2m_3\Delta  
\cr\cr &\displaystyle  
  \left(\frac{f_{1T}^{\frac{3}{2}\frac{3}{2}1}}{m_1^2} 
      +\frac{f_{2T}^{\frac{3}{2}\frac{3}{2}1}}{m_2^2}   
      +\left(1-\frac{p_1\!\cdot\!p_2}{2m_1m_2}\right)
        \frac{f_{3T}^{\frac{3}{2}\frac{3}{2}1}}{m_3^2}
      -\frac{\Delta^2}{4m_1^2m_2^2m_3^2} f_L^{\frac{3}{2}\frac{3}{2}1}
\right),\end{array}\end{array}\]
and
\[
\begin{array}{l}\displaystyle 
 \frac{1}{2}(M_{3/2,1}-M_{-3/2,-1})=i\sqrt{-\frac{1}{3}(m_1m_2+p_1\cdot p_2)}\;
 \Delta\left(\frac{p_2\cdot p_3}{m_1} g_{1T}^{\frac{3}{2}\frac{3}{2}1}
            +g_M^{\frac{3}{2}\frac{3}{2}1}\right),
\cr\cr\displaystyle 
 \frac{1}{2}(M_{1/2,-1}-M_{-1/2,1})=i\sqrt{-\frac{1}{3}(m_1m_2+p_1\cdot p_2)}\;
 \Delta\left(\frac{p_1\cdot p_3}{m_2} g_{2T}^{\frac{3}{2}\frac{3}{2}1}
            -g_M^{\frac{3}{2}\frac{3}{2}1}\right),
\cr\cr\displaystyle 
 \frac{1}{2}(M_{1/2,1}-M_{-1/2,-1})=i\sqrt{-\frac{1}{9}(m_1m_2+p_1\cdot p_2)}\;
 \Delta\!\left(-\frac{p_2\!\cdot\!p_3}{m_1}g_{1T}^{\frac{3}{2}\frac{3}{2}1}\!
              +\!\frac{p_1\!\cdot\! p_3}{m_2}g_{2T}^{\frac{3}{2}\frac{3}{2}1}
              \!-\!2\frac{p_1\!\cdot\! p_2}{m_1m_2}
              g_M^{\frac{3}{2}\frac{3}{2}1}\right),
\cr\cr\displaystyle 
\frac{1}{2}(M_{3/2,0}-M_{-3/2,0})=
 -i\sqrt{-2(m_1m_2+p_1\cdot p_2)}\; \frac{\Delta}{2m_3} p_1\cdot p_2
  g_{3T}^{\frac{3}{2}\frac{3}{2}1},
\cr\cr\displaystyle  
\begin{array}{rl}\displaystyle  
 \frac{1}{2}(M_{1/2,0}-M_{-1/2,0})& \displaystyle  
=i\sqrt{-\frac{2}{9}(m_1m_2+p_1\cdot p_2)}\;  m_1m_2m_3\Delta  
\cr\cr &\displaystyle  
  \left(\frac{g_{1T}^{\frac{3}{2}\frac{3}{2}1}}{m_1^2} 
      +\frac{g_{2T}^{\frac{3}{2}\frac{3}{2}1}}{m_2^2}   
      +\left(1+\frac{p_1\!\cdot\! p_2}{2m_1m_2}\right)
        \frac{g_{3T}^{\frac{3}{2}\frac{3}{2}1}}{m_3^2}
      -\frac{\Delta^2}{4m_1^2m_2^2m_3^2} g_L^{\frac{3}{2}\frac{3}{2}1}
\right).\end{array}\end{array}\]

The exclusive decay rates for all of these processes can be found via the 
helicity amplitudes of course and, by comparison with the rates found in the 
covariant formalism, this provides an independent check on the correctness of 
the expressions connecting the two formalisms. It also allows us to deduce
the density matrices, if needed, without recourse to further tedious
evaluation of traces incorporating polarization projectors. It is worth
remebering that to each helicity amplitude is associated a particular
angular decay contribution.

\section{Meson decays}

\subsection{Covariant amplitudes}

\noindent
$\underline{0 \rightarrow 0 + 0}$

This is perfectly trivial. One only has 
$${\cal L}^{000}=g^{000},$$ 
leading to the decay width
\[
\Gamma^{000}=\frac{\Delta}{16\pi m_1^3}|g^{000}|^2.
\]

Examples are the parity-conserving decay $\sigma \rightarrow \pi\pi$ and
parity-violating decays like $K \rightarrow \pi\pi$.

\noindent
$\underline{0 \rightarrow 0 + 1}$

This too is effectively unique. We take
\[
{\cal L}^{001}=\frac{1}{2}(p_2-p_1)_\nu\varepsilon^{\nu *}_3 g^{001},
\]
whereupon the resulting width equals
\[
\Gamma^{001}=\frac{\Delta^3}{64\pi m_1^3m_3^2}| g^{001}|^2. 
\]

The processes $D\rightarrow \bar{K}^*\pi,\quad \phi\pi,\quad\bar{K}^*K$
exemplify this case.

\noindent
$\underline{0 \rightarrow 1/2 + 1/2}$

This can be obtained from the decay $1/2 \rightarrow 1/2 + 0$ by interchanging
particles 1 and 3, since the Lagrangian here reads,
\[
{\cal L}^{0\frac{1}{2}\frac{1}{2}} = \bar{u}_2[f^{0\frac{1}{2}\frac{1}{2}} +
                                   g^{0\frac{1}{2}\frac{1}{2}}\gamma_5]v_3
\]
and by recalling the connection between $v$ and $u$, spelled out in the
appendix. This implies that the sum over spins is obtainable by interchanging
the labels 1 and 3, remembering also to multiply the conjugation factor
$(-1)^{2j}$. Upon including the correct phase space factor, this gives,
\[
\Gamma^{0\frac{1}{2}\frac{1}{2}} =\frac{\Delta}{4\pi m_1^3}  
[(p_2\cdot p_3-m_2m_3)|f^{0\frac{1}{2}\frac{1}{2}}|^2  +
 (p_2\cdot p_3+m_2m_3)|g^{0\frac{1}{2}\frac{1}{2}}|^2]. 
\]
One may safely apply this substitution rule to save repetitious calculations 
of spin sums in other instances. Hereafter we shall do this and suppress a
host of similar results which can be read off from others.

Hadronic examples of these decays are very unusual. A rarity is $\eta_c
\rightarrow N\bar{N}.$

\noindent
$\underline{0 \rightarrow 1 + 1}$

This time we encounter three independent amplitudes if we make no assumptions
about parity. It corresponds to the number of ways that two spin 1 systems
can be added. We write the effective Lagrangian in the form,
\[
{\cal L}^{011}=\varepsilon^{\mu *}_2\varepsilon^{\nu *}_3
[g_T^{001} (g_{\mu\nu}p_2\cdot p_3-p_{2\nu}p_{3\mu})+g_L^{011} p_{2\nu}p_{3\mu}
+ g_M^{011}\epsilon_{\mu\nu\alpha\beta}p_2^\alpha p_3^\beta]. 
\]
Summing over final spins, it is straightforward to work out the decay width,  
\[
\Gamma^{011}=
\frac{\Delta}{32\pi m_1^3}\left[(\Delta^2+6m_2^2m_3^2)|g_T^{011}|^2 + 
\Delta^2 |g_M^{011}|^2 + \frac{\Delta^4}{8m_2^2 m_3^2}|g_L^{011}|^2 +
\Delta^2{\rm Re}(g_T^{011}g_L^{011*}) \right]. 
\]
If the decay is radiative, one of the final particles is a photon (say 3) and
then gauge invariance requires $g_L^{011}\rightarrow 0$; the width reduces to
\[
\Gamma^{011} \rightarrow \frac{\Delta^3}{32\pi m_1^3}
\left(|g_T^{011}|^2+|g_M^{011}|^2\right),
\qquad \Delta \rightarrow m_1^2-m_2^2. 
\]
(Actually, one or other of the two couplings $g_M, g_T$ must be set to zero,
since electromagnetism conserves parity.) It is even possible for both final 
particles to be photons, as in $\pi^0$ decay, in which case $\Delta$ 
simplifies further to $m_1^2$.

The decays $K\rightarrow \gamma\gamma, \quad \eta_c\rightarrow \rho\rho,\quad
D\rightarrow K^{*}\rho$ typify this case.

\noindent
$\underline{0 \rightarrow 1/2 + 3/2}$

This case is obtainable $3/2 \rightarrow 1/2 + 0$ by substitution; remember
to adjust the external spin average by a factor of 4 for the partial decay 
width.

No examples of this process stand out in the Particle Data tables.

\noindent
$\underline{0 \rightarrow 3/2 + 3/2}$

Again we obtain this from $3/2 \rightarrow 3/2 + 0$ by substitution and
insert an extra factor of 4 to make up for the spin average.

It will be a long time before such exclusive modes are detected.

\noindent
$\underline{1 \rightarrow 0 + 0}$

The effective Lagrangian here is just a labelling permutation 
($1 \leftrightarrow 3$) of the previous process $0\rightarrow 0 + 1$; 
thus we get 
\[
\Gamma^{100}=\frac{\Delta^3}{192\pi m_1^5}|g^{100}|^2, 
\]
where an extra factor 1/3 accounts for the average over the spin states of
the initial vector meson. The decays $\rho \rightarrow \pi\pi, \omega
\rightarrow \pi\pi$ are typical.

\noindent
$\underline{1 \rightarrow 0 + 1}$

The effective Lagrangian is obtainable from the earlier process,
$0 \rightarrow 1 + 1$, by permuting the labels 1 and 2. Hence the decay is
found to be
\[
\Gamma^{101}=
\frac{\Delta}{96\pi m_1^3} \left[(\Delta^2+ 6m_1^2m_3^2)|g_T^{101}|^2 + 
\Delta^2 |g_M^{101}|^2 + \frac{\Delta^4}{8m_1^2 m_3^2}|g_L^{101}|^2+
\Delta^2 {\rm Re}(g_T^{101}g_L^{101*}) \right]. 
\]
In particular, a radiative vector meson decay into a scalar meson requires 
$g_L^{101}\rightarrow 0$ in which case the width reduces to
\[
\Gamma^{101}\rightarrow\frac{\Delta^3}{96\pi m_1^3}
(|g_T^{101}|^2  + |g_M^{101}|^2), 
\]
with one of $g_M$ or $g_T$ vanishing by parity conservation.
Examples of such vector decays are $\psi \rightarrow \omega \pi,\quad 
D^*\rightarrow D\gamma.$

Here we can point to $\phi\rightarrow\rho\pi,\quad B^*\rightarrow B \gamma$
as typical cases.

\noindent
$\underline{1 \rightarrow 1/2 + 1/2}$

This situation lends itself to the substitution rule, and apart from the spin 
averaging factor (namely a factor of 2/3), can be derived directly from
$1/2 \rightarrow 1/2 + 1$. It is
\[
 \Gamma^{1\frac{1}{2}\frac{1}{2}}\!=\!\frac{\Delta^3}{48\pi m_1^3} \left[
 (m_2m_3\!-\!p_2\!\cdot\!p_3)
 \left(\frac{|f_E^{\frac{1}{2}\frac{1}{2}1}|^2}{m_1^2}\!+
 2|f_M^{\frac{1}{2}\frac{1}{2}1}|^2 \right)\! - \!
 (m_2m_3\!+\!p_2\!\cdot\!p_3) 
 \left(\frac{|g_E^{\frac{1}{2}\frac{1}{2}1}|^2}{m_1^2}\!+
 2|g_M^{\frac{1}{2}\frac{1}{2}1}|^2\right)\! \right] 
\]

Typical examples are $\psi\rightarrow N\bar{N}, \Sigma\bar{\Sigma}.$

\noindent
$\underline{1 \rightarrow 1 + 1}$

This is the most complicated purely mesonic process which we consider.
It brings in seven independent amplitudes in general, no assumption being made
about parity. The most elegant decomposition of the effective Lagrangian is 
\[
{\cal L}^{111}=\varepsilon^{\mu *}_2\varepsilon^{\nu *}_3
              {\cal M}_{\lambda\mu\nu}\varepsilon_1^\lambda, 
\]
with 
\[
\begin{array}{rl}
 {\cal M}_{\lambda\mu\nu}= & \frac{1}{2}(p_3-p_2)_\lambda
 [g_{1T}^{111}(p_2\cdot p_3g_{\mu\nu}-p_{3\mu}p_{2\nu})+
  g_{1M}^{111}\epsilon_{\mu\nu\alpha\beta}p_2^\alpha p_3^\beta]\quad + \cr
 & \frac{1}{2}(p_1-p_3)_\mu
 [g_{2T}^{111}(p_3\cdot p_1 g_{\nu\lambda}-p_{1\nu}p_{3\lambda})+
  g_{2M}^{111}\epsilon_{\nu\lambda\alpha\beta}p_3^\alpha p_1^\beta]\quad + \cr 
 & \frac{1}{2}(p_2-p_1)_\nu
 [g_{3T}^{111}(p_1\cdot p_2 g_{\lambda\mu}-p_{2\lambda}p_{1\mu})+
  g_{3M}^{111}\epsilon_{\lambda\mu\alpha\beta}p_1^\alpha p_2^\beta]\quad + \cr 
 & \frac{1}{8}g_L^{111} (p_3-p_2)_\lambda(p_1-p_3)_\mu(p_2-p_1)_\nu.
\end{array}
\]
After a little work the width of the process is calculated to be 
\[
\begin{array}{rl}
\Gamma=\frac{\Delta^3}{192\pi m_1^3} & \displaystyle\left\{ \frac{1}{2m_1^2}
 [(\Delta^2+6m_2^2m_3^2)|g_{1T}^{111}|^2+\Delta^2|g_{1M}^{111}|^2] 
 +\frac{\Delta^4}{16m_1^2m_2^2m_3^2}\vert g_L^{111}\vert^2\right. + \cr & 
 \displaystyle\frac{1}{2m_2^2}[(\Delta^2+6m_1^2m_3^2)|g_{2T}^{111}|^2 
 + \Delta^2|g_{2M}^{111}|^2]\; + \cr 
 & \displaystyle\frac{1}{2m_3^2}[(\Delta^2+6m_1^2m_2^2)|g_{3T}^{111}|^2 
 +\Delta^2|g_{3M}^{111}|^2]\; + \cr 
 & \left. \displaystyle 
 2{\rm Re}(m_1^2 g_{2T}^{111}g_{3T}^{111*}\!+\!m_2^2g_{1T}^{111}g_{3T}^{111*}
           \!+\!m_3^2g_{1T}^{111}g_{2T}^{111*})
 \!+\!\!\frac{\Delta^2}{2}{\rm Re}[(\frac{g_{1T}^{111}}{m_1^2}
                          \!+\!\frac{g_{2T}^{111}}{m_2^2}
                          \!+\!\frac{g_{3T}^{111}}{m_3^2})g_L^{111*}\!\right\}
\end{array}
\]
For the particular case that one of the final particles is a photon, say 2,
gauge invariance on that leg, $p_2^\mu{\cal M}_{\lambda\mu\nu} =0$,
results in three conditions, 
\[
g_L^{111}\rightarrow 0, \quad g_{2T}^{111}\rightarrow0, 
\quad g_{2M}^{111}\rightarrow 0, 
\]
reducing number of amplitudes for $1\rightarrow \gamma 1$ to four (which we 
check presently via the helicity formalism); the corresponding decay width 
reads 
\[
\Gamma^{111} \rightarrow \frac{\Delta^5}{384\pi m_1^3} \left[ 
\frac{1}{m_1^2}(|g_{1T}^{111}|^2+|g_{1M}^{111}|^2)  +
\frac{1}{m_3^2}(|g_{3T}^{111}|^2+|g_{3M}^{111}|^2) \right], \qquad
\Delta\rightarrow m_1^2-m_3^2.
\]
When both final state particles are photons so that gauge invariance applies 
to leg 3 as well, only $g_{1T}^{111}$ and $g_{1M}^{111}$ survive and one
remains with
\[
 \Gamma^{111}\rightarrow\frac{m_1^5}{384\pi}
 (|g_{1T}^{111}|^2+|g_{1M}^{111}|^2).
\]
Examples of such decay processes are $\psi \rightarrow K^*\bar{K}^*, \psi
\rightarrow \gamma\gamma.$

\noindent
$\underline{1 \rightarrow 1/2 + 3/2}$

We get this from $3/2\rightarrow 1/2+1$ by substitution and shall therefore 
not repeat the expression apart from reminding the reader about the initial 
spin factor in the partial decay width.

Instances of such decays are rare, but $\psi\rightarrow\bar{\Xi}\Xi^*$
is a case in point.

\noindent
$\underline{1 \rightarrow 3/2 + 3/2}$

Substitution allows us to deduce this from $3/2 \rightarrow 3/2 + 1$.

Two examples are $\psi\rightarrow\Delta\bar{\Delta},\Xi^*\bar{\Xi}^*.$

\subsection{Helicity amplitudes}

\noindent
$\underline {0\rightarrow 0 + 0}$

No analysis is required here.
\[M_{0,0} = g^{000}.\]

\noindent
$\underline {0\rightarrow 0 + 1}$

This case is equally simple:
\[M_{0,0}=\frac{\Delta}{2m_3}g^{001}.\]

\noindent
$\underline{0 \rightarrow 1/2 + 1/2}$

When contracting the effective Lagrangian (a crossed version of
$1/2 \rightarrow 1/2 + 0$) against the spinors,
\[
{\cal L} = \bar{u}^{(\lambda_2)}(p_2){\cal M}v^{(\lambda_3)}(p_3),
\]
it is worth pointing out that, here and below, one runs across kinematic
factors which can be simplified as follows:
\[
 \sqrt{(E_3-m_3)(E_2\pm m_2)} + \sqrt{(E_3+m_3)(E_2\mp m_2)} =
 \sqrt{2(p_2\cdot p_3 \mp m_2m_3)}.
\]
In this way, for this case, one arrives at the helicity amplitudes,
\[\frac{1}{2}
(M_{1/2,1/2}+M_{-1/2,-1/2})=-\sqrt{2p_2\cdot p_3 - 2m_2m_3}\; 
                           f^{0\frac{1}{2}\frac{1}{2}},
\]
\[\frac{1}{2}
(M_{1/2,1/2}-M_{-1/2,-1/2})=-i\sqrt{2p_2\cdot p_3 + 2m_2m_3}\; 
                           g^{0\frac{1}{2}\frac{1}{2}}.
\]

\noindent
$\underline {0\rightarrow 1 + 1}$ 
\[
M_{0,0}=m_2m_3 g_T^{011} +\frac{\Delta^2}{4m_2m_3} g_L^{011} .
\]
The parity symmetric combinations under final helicity reversal are 
\[\frac{1}{2}
(M_{1,1}+M_{-1,-1})=p_2\cdot p_3 g_T^{011},\quad 
M_{1,1}-M_{-1,-1}=i\Delta g_M^{011} 
\]

\noindent
$\underline{0 \rightarrow 1/2 + 3/2}$

The contraction here brings in the spin 3/2 antiparticle wavefunctions
which are obtained from the direct product topmost state, $v_\nu^{(3/2)}(p_3)
\equiv\varepsilon_\nu^{(1)*}(p_3)v^{(1/2)}(p_3)$ by spin lowering. As with
the decay $0\rightarrow 1/2 + 1/2$, we are restricted to helicity values
$\lambda_3=\lambda_2=\pm 1/2$ because of the initial spin. We thereby obtain
\[\frac{1}{2}
(M_{1/2,1/2}+M_{-1/2,-1/2})=-\sqrt{\frac{1}{3}(p_2\cdot p_3 - m_2m_3)}\;
                           \frac{\Delta}{m_3}f^{0\frac{1}{2}\frac{3}{2}},
\]
\[\frac{1}{2}
(M_{1/2,1/2}-M_{-1/2,-1/2})=-i\sqrt{\frac{1}{3}(p_2\cdot p_3 + m_2m_3)}\;
                           \frac{\Delta}{m_3}g^{0\frac{1}{2}\frac{3}{2}}.
\]
Observe how this represents a crossed version of $3/2 \rightarrow 1/2+0$.

\noindent
$\underline{0 \rightarrow 3/2 + 3/2}$

There occur four helicity amplitudes now, because $\lambda_3=\lambda_2$ can
run through all values. The results read,
\[\frac{1}{2}
(M_{3/2,3/2}+M_{-3/2,-3/2})=-\sqrt{2(p_2\cdot p_3 - m_2m_3)}\;
                           p_2\cdot p_3f_T^{0\frac{3}{2}\frac{3}{2}},
\]
\[\frac{1}{2}
(M_{3/2,3/2}-M_{-3/2,-3/2})=-i\sqrt{2(p_2\cdot p_3 + m_2m_3)}\;
                            p_2\cdot p_3g_T^{0\frac{3}{2}\frac{3}{2}},
\]
\[\frac{1}{2}
(M_{1/2,1/2}+M_{-1/2,-1/2})=-\sqrt{\frac{2}{9}(p_2\cdot p_3 - m_2m_3)}
            \left[(p_2\cdot p_3+2m_2m_3)f_T^{0\frac{3}{2}\frac{3}{2}}
           + \frac{\Delta^2}{2m_2m_3}f_L^{0\frac{3}{2}\frac{3}{2}}\right],
\]
\[\frac{1}{2}
(M_{1/2,1/2}-M_{-1/2,-1/2})=i\sqrt{\frac{2}{3}(p_2\cdot p_3 + m_2m_3)}
            \left[(p_2\cdot p_3-2m_2m_3)g_T^{0\frac{3}{2}\frac{3}{2}}
           -\frac{\Delta^2}{2m_2m_3}g_L^{0\frac{3}{2}\frac{3}{2}}\right].
\]

\noindent
$\underline {1\rightarrow 0 + 0}$

This is a crossed version of an earlier case.
\[M_{0,0} = \frac{\Delta}{2m_1}g^{100}.\]

\noindent
$\underline {1\rightarrow 0 + 1}$ 
\[
M_{0,0}=-m_1m_3 g_T^{101} -\frac{\Delta^2}{4m_1m_3} g_L^{101} 
\]
\[\frac{1}{2}(M_{0,1}+M_{0,-1})=-p_1\cdot p_3 g_T^{101},\quad 
\frac{1}{2}(M_{0,1}-M_{0,-1})=i\frac{\Delta}{2} g_M^{101}.
\]

\noindent
$\underline{1 \rightarrow 1/2 + 1/2}$

In terms of the covariant electric and magnetic decomposition,
\[\frac{1}{2}
(M_{1/2,1/2}+M_{-1/2,-1/2})=-\sqrt{2(p_2\cdot p_3 - m_2m_3)}\;
                             \frac{\Delta}{2m_1}f_E^{1\frac{1}{2}\frac{1}{2}},
\]
\[\frac{1}{2}
(M_{1/2,1/2}-M_{-1/2,-1/2})=-i\sqrt{2(p_2\cdot p_3+m_2m_3)}\;
                              \frac{\Delta}{2m_1}g_E^{1\frac{1}{2}\frac{1}{2}},
\]
\[\frac{1}{2}
(M_{1/2,-1/2}+M_{-1/2,1/2})=\sqrt{2(p_2\cdot p_3 - m_2m_3)}\;
                            \frac{\Delta}{2}f_M^{1\frac{1}{2}\frac{1}{2}},
\]
\[\frac{1}{2}
(M_{1/2,1/2}-M_{-1/2,-1/2})=i\sqrt{2(p_2\cdot p_3 + m_2m_3)}\;
                             \frac{\Delta}{2}g_M^{1\frac{1}{2}\frac{1}{2}}.
\]

\noindent
$\underline {1\rightarrow 1 + 1}$

A total of seven amplitudes out of a possible nine survive because the
amplitudes $M_{1,-1}, M_{-1,1}$ are excluded by the initial spin magnitude.
This agrees of course with other counting methods.
\[
M_{0,0} =-\frac{\Delta}{2}m_1m_2 m_3
           \left(\frac{g_{1T}^{111}}{m_1^2}  +
                 \frac{g_{2T}^{111}}{m_2^2}  +
                 \frac{g_{3T}^{111}}{m_3^2}  + 
\frac{\Delta^2}{4m_1^2m_2^2m_3^3}g_L^{111} \right). 
\]
Well-defined parity combinations are 
\[
M_{1,1}+M_{-1,-1} =\frac{\Delta}{m_1}p_2\cdot p_3 g_{1T}^{111},\quad
M_{1,1}-M_{-1,-1} =i\frac{\Delta^2}{2m_1} g_{1M}^{111}, 
\]
\[
M_{0,1}+M_{0,-1} =i\frac{\Delta^2}{2m_2} g_{2M}^{111},\quad
M_{0,1}-M_{0,-1} =\frac{\Delta}{m_2}p_2\cdot p_3 g_{2T}^{111},
\]
\[
M_{1,0}+M_{-1,0} =i\frac{\Delta^2}{2m_3} g_{3M}^{111}, \quad
M_{1,0}-M_{-1,0} =-\frac{\Delta}{m_3}p_2\cdot p_3 g_{3T}^{111}. 
\]
We can see directly from above that if particle 2 is a photon,
the couplings $g_L^{111}, g_{2T}^{111}, g_{2M}^{111}$, must disappear,
like we established before (and $g_{3T}^{111} = g_{3M}^{111} \rightarrow 0$
if the third leg is electromagnetic too).

\noindent
$\underline{1 \rightarrow 1/2 + 3/2}$
\[\begin{array}{l}\displaystyle
\frac{1}{2}(M_{1/2,3/2}+M_{-1/2,-3/2}) =\sqrt{2(p_2\cdot p_3-m_2m_3)}\;
p_1\cdot p_3 f_T^{1\frac{1}{2}\frac{3}{2}},
\cr\cr\displaystyle
\frac{1}{2}(M_{1/2,1/2}+M_{-1/2,-1/2}) =
\sqrt{\frac{1}{3}(p_2\cdot p_3-m_2m_3)}\;
\left(2m_1m_3 f_{T}^{1\frac{1}{2}\frac{3}{2}}
     +\frac{\Delta^2}{2m_1m_3}f_L^{1\frac{1}{2}\frac{3}{2}}\right),
\cr\cr\displaystyle
\frac{1}{2}(M_{1/2,-1/2}+M_{-1/2,1/2}) =
\sqrt{\frac{1}{3}(p_2\cdot p_3-m_2m_3)}\;
\left(p_1\cdot p_3f_{T}^{1\frac{1}{2}\frac{3}{2}}
     +\frac{\Delta^2}{2m_3}f_M^{1\frac{1}{2}\frac{3}{2}}\right),
\cr\cr\displaystyle
\frac{1}{2}(M_{1/2,3/2}-M_{-1/2,-3/2}) =i\sqrt{2(p_2\cdot p_3+m_2m_3)}\;
p_1\cdot p_3 g_T^{1\frac{1}{2}\frac{3}{2}}
\cr\cr\displaystyle
\frac{1}{2}(M_{1/2,1/2}-M_{-1/2,-1/2}) =i
\sqrt{\frac{1}{3}(p_2\cdot p_3+m_2m_3)}\;
\left(2m_1m_3 g_{T}^{1\frac{1}{2}\frac{3}{2}}
     +\frac{\Delta^2}{2m_1m_3}g_L^{1\frac{1}{2}\frac{3}{2}}\right),
\cr\cr\displaystyle
\frac{1}{2}(M_{1/2,-1/2}-M_{-1/2,1/2}) =i
\sqrt{\frac{1}{3}(p_2\cdot p_3+m_2m_3)}\;
\left(p_1\cdot p_3g_{T}^{1\frac{1}{2}\frac{3}{2}}
     +\frac{\Delta^2}{2m_3}g_M^{1\frac{1}{2}\frac{3}{2}}\right).
\end{array}\]

\noindent
$\underline{1 \rightarrow 3/2 + 3/2}$
\[\begin{array}{l}\displaystyle
\frac{1}{2}(M_{3/2,3/2}+M_{-3/2,-3/2}) =-\sqrt{2(p_2\cdot p_3-m_2m_3)}\;
\frac{\Delta}{2m_1}p_2\cdot p_3 f_{1T}^{1\frac{3}{2}\frac{3}{2}},
\cr\cr\displaystyle
\frac{1}{2}(M_{3/2,1/2}+M_{-3/2,-1/2})=-\sqrt{\frac{1}{3}(p_2\cdot p_3-m_2m_3)}
\;\Delta\left(\frac{p_1\cdot p_2}{m_3} f_{3T}^{1\frac{3}{2}\frac{3}{2}}
             -f_M^{1\frac{3}{2}\frac{3}{2}}\right),
\cr\cr\displaystyle
\frac{1}{2}(M_{1/2,3/2}+M_{-1/2,-3/2})=-\sqrt{\frac{1}{3}(p_2\cdot p_3-m_2m_3)}
\;\Delta\left(\frac{p_1\cdot p_3}{m_2} f_{2T}^{1\frac{3}{2}\frac{3}{2}}
             -f_M^{1\frac{3}{2}\frac{3}{2}}\right),
\cr\cr\displaystyle
\frac{1}{2}(M_{1/2,-1/2}+M_{-1/2,1/2})=-\sqrt{\frac{1}{9}(p_2\cdot p_3-m_2m_3)}
\;\Delta\left(\frac{p_1\cdot p_3}{m_3} f_{2T}^{1\frac{3}{2}\frac{3}{2}}
             -\frac{p_1\cdot p_2}{m_3} f_{3T}^{1\frac{3}{2}\frac{3}{2}}
             +f_M^{1\frac{3}{2}\frac{3}{2}}\right),
\cr\cr\displaystyle
\begin{array}{rl}
\frac{1}{2}(M_{1/2,1/2}+M_{1/2,1/2})=&\displaystyle
-\sqrt{\frac{1}{9}(p_2\cdot p_3-m_2m_3)}\;m_1m_2m_3\Delta
\cr &\displaystyle
 \left[\left(1+\frac{p_2\cdotp_3}{2m_2m_3}\right)
                            \frac{f_{1T}^{1\frac{3}{2}\frac{3}{2}}}{m_1^2}
                           +\frac{f_{2T}^{1\frac{3}{2}\frac{3}{2}}}{m_2^2}
                           +\frac{f_{3T}^{1\frac{3}{2}\frac{3}{2}}}{m_3^2}
                           -\frac{\Delta^2}{4m_1^2m_2^2m_3^2}
                             f_L^{1\frac{3}{2}\frac{3}{2}}\right]
\end{array}
\cr\cr\displaystyle
\frac{1}{2}(M_{3/2,3/2}-M_{-3/2,-3/2}) =-i\sqrt{2(p_2\cdot p_3+m_2m_3)}\;
\frac{\Delta}{2m_1}p_2\cdot p_3 g_{1T}^{1\frac{3}{2}\frac{3}{2}}
\cr\cr\displaystyle
\frac{1}{2}(M_{3/2,1/2}-M_{-3/2,-1/2})=\sqrt{\frac{1}{3}(p_2\cdot p_3+m_2m_3)}
\;\Delta\left(\frac{p_1\cdot p_2}{m_3} g_{3T}^{1\frac{3}{2}\frac{3}{2}}
             -g_M^{1\frac{3}{2}\frac{3}{2}}\right),
\cr\cr\displaystyle
\frac{1}{2}(M_{1/2,3/2}-M_{-1/2,-3/2})=-i\sqrt{\frac{1}{3}(p_2\cdot p_3+m_2m_3)}
\;\Delta\left(\frac{p_1\cdot p_3}{m_2} g_{2T}^{1\frac{3}{2}\frac{3}{2}}
             +g_M^{1\frac{3}{2}\frac{3}{2}}\right),
\cr\cr\displaystyle
\frac{1}{2}(M_{1/2,-1/2}-M_{-1/2,1/2})=-i\sqrt{\frac{1}{3}
    (p_2\!\cdot\! p_3\!+\!m_2m_3)}
\;\Delta\left(\frac{p_1\!\cdot\!p_3}{m_2}g_{2T}^{1\frac{3}{2}\frac{3}{2}}
           \!+\!\frac{p_1\!\cdot\!p_2}{m_3}g_{3T}^{1\frac{3}{2}\frac{3}{2}}
           \!+\!\frac{2p_2\!\cdot\!p_3}{m_2m_3}g_M^{1\frac{3}{2}\frac{3}{2}}\right),
\cr\cr\displaystyle
\begin{array}{rl}
\frac{1}{2}(M_{1/2,1/2}-M_{1/2,1/2})\!=\! &\displaystyle
-i\sqrt{\frac{1}{9}(p_2\cdot p_3+m_2m_3)}\;m_1m_2m_3\Delta
\cr & \displaystyle
\left[\left(1-\frac{p_2\cdotp_3}{2m_2m_3}\right)
                            \frac{g_{1T}^{1\frac{3}{2}\frac{3}{2}}}{m_1^2}
                           +\frac{g_{2T}^{1\frac{3}{2}\frac{3}{2}}}{m_2^2}
                           +\frac{g_{3T}^{1\frac{3}{2}\frac{3}{2}}}{m_3^2}
                           -\frac{\Delta^2}{4m_1^2m_2^2m_3^2}
                             g_L^{1\frac{3}{2}\frac{3}{2}}\right]
\end{array}\end{array}\]

In all the cases above one may readily verify that the decay width 
expressions determined previously emerge upon summing over helicity
states; this is a helpful independent check.

\section{Magnitudes}

The Particle Data Tables contain vast quantities of information about
decay processes \cite{PD}, all amenable to theoretical analysis. 
Examining every single channel (even for a single particle like $\psi$), 
in order to test theoretical predictions, represents a major undertaking.
Many publications have appeared which focus on one aspect or another of
such decays, such as semileptonic channels or nonleptonic channels
involving a double change in quantum numbers. In this paper, we have
chosen instead to concentrate on amplitude analysis. For the rest,
we shall content ourselves with looking at some typical electromagnetic,
weak and strong decays in order to make an important point: that it rather
easy to derive the correct partial decay width, within a factor of about
two ($\sim$ a Clebsch-Gordan coefficient), just by incorporating the proper
phase space, CKM angles and invoking the magnitudes,
$$ g_{\rm strong} \sim 10, \quad e \simeq 0.3, \quad
   G_Fm_p^2/2\sqrt{2} = 3.62 \times 10^{-6}.$$
Our main conclusion is that a stringent test of theoretical predictions
demands that experimental fits should be good to within 10\% at least; 
otherwise there is little point in extolling the virtues of one
theoretical model over another.

To substantiate this statement, look at weak processes $0^-\rightarrow 
0^-0^-$ and $0^-\rightarrow 0^- 1^-$, which are governed by a single 
coupling constant, whose magnitude can be directly obtained
from the decay width. Because, as defined, $g^{000}$ has dimensions, we shall
rescale it to a dimensionless number by dividing it by the mass of the
decaying particle--- this seems the most natural choice.
($g^{001}$ is automatically dimensionless and does not need massaging.)
For $B$-meson decays, we further divide out the CKM matrix elements,
$\vert V_{cb}\vert$ and $\vert V_{ud}\vert$ or $\vert V_{cs}\vert$,
which we anticipate to be indirectly present from the current-current
picture model; so for example, one should understand that
$g^{BD\pi} = g^{000}(B\rightarrow D\pi)/m_BV_{cb}V_{ud}$ below, etc.
The results for typical $B$-meson decays read (in units of $10^{-6}$)
\[\begin{array}{l}
g^{B^+D^0\pi^+}=3.84 (0.12)   \quad  g^{B^0D^-\pi^+}=2.94 (0.14) \cr

g^{B^+D^{*0}\pi^+}=3.42 (0.15) \quad g^{B^0 D^{*-}\pi^+}=2.46 (0.15) \cr

g^{B^+\psi\pi^+}=3.15(0.42) \cr

g^{B^+\psi K^+}=3.48(0.32), \quad g^{B^0\psi K^0}=3.06(0.39) \cr

g^{B^+D^0\rho^+}=2.13 (0.14)   \quad g^{B^0D^-\rho^+}=1.66 (0.16),
\end{array}\]
where the entries in the parenthesis are relative errors derived from
the uncertainties of the measured branching ratios and the elements
of the Cabibbo-Kobayashi-Maskawa matrix. We see that the couplings 
$g^{BD\pi}$along with $g^{B\psi\pi}$ and $g^{B\psi K}$ hover around 3,
while the couplings $g^{BD\rho}$ are somewhat smaller. These values are
comparable to $m_p^2G_F/2\sqrt{2}$.

With respect to strong processes, parity is conserved and there are
fewer independent couplings as a result. From the data one determines
the following typical coupling constants,
\[\begin{array}{l}
g^{K^*K\pi}=11.2 \cr
f^{\Delta N \pi}=10.9   \cr    g^{N^*N\pi}=9.65, \quad N^*= N(1440).
\end{array}\]
Here, $f^{\frac{3}{2}\frac{1}{2}0}$ has been rescaled by a factor
$m_\Delta$ while $g^{\frac{1}{2}\frac{1}{2}0}$ is already dimensionless.
These numbers are typical of strong couplings, as we have asserted above.

Coupling constants for electromagnetic processes of ground state
hadrons accord well\footnote{In this connection we note that relativistic
spin-flavour symmetry produces a particular invariant amplitude, which
can be re-expressed in terms our previous amplitudes, via (with $m_3=0$),\\
$(m_1-m_2)^2\varepsilon^{*3\nu}\bar{u}_2
\epsilon_{\nu\alpha\beta\lambda}p_1^\alpha p_2^\beta u_1^\lambda =
\varepsilon^{*\nu}\bar{u}_2[(p_1\cdot p_3 g_{\lambda\nu}-p_{3\lambda}p_{1\nu})
\gamma_5(m_1^2-m_2^2) - m_1(p_3-p_2)_\lambda\epsilon_{\nu\alpha\beta\rho}
p_1^\alpha p_2^\beta \gamma^\rho]u_1^\lambda,$} with supermultiplet
predictions \cite{RDDL}, so it is of interest to look for the effects
connected with excitations; in particular the coupling,
\[f_M^{N^*p\gamma}=0.0988, \] 
which has been rescaled by a factor of $m_{N^*}^2$, is smaller than
electromagnetic value magnitude $e$; this is not
surprising because it represents a transition between an excited and a
ground state---it should therefore be smaller than an electromagnetic
transition element between $\ell=0$ hadron states. In fact the larger the
excitation number difference (and hence the baryon mass values) the smaller 
do we anticipate the coupling to be, so all is crudely in order here.

In summary, we suggest that all decay analyses, including nonleptonic weak
decays, be converted into the covariant decompositions prescribed in the
text, for reasons of economy and elegance, and without assuming factorization. 
As well, we believe that theoretical models (that can be made to accord 
with data all too readily) will be better differentiated by more accurate 
experimental results; at the moment there is little to choose between them.

\appendix
\section{Notation}

We use a (+,-,-,-) Minkowskian metric $g$ and the state normalization
\[
2\pi\delta(p^2-m^2)\langle p'j'\lambda'|pj\lambda\rangle = (2\pi)^4
\delta^4(p'-p)\delta_{j'j}\delta_{\lambda\lambda'}\]
or
\[
\langle{\bf p'}j'\lambda'|{\bf p}j\lambda\rangle = 2p_0
(2\pi)^3\delta^3({\bf p'}-{\bf p})\delta_{jj'}\delta_{\lambda'\lambda}.
\]
We shall denote the intrinsic parity of a state by $\eta$, as determined by
strong interactions.  A two-body decay process of the type $1\rightarrow 2+3$
is viewed with all momenta flowing outwards (to ensure that any symmetry
and crossing properties are readily implemented); thus to an incoming decaying
particle 1 is ascribed a momentum -$p_1$, whereas the outgoing momenta are
simply written $p_2$ and $p_3$. While it may seem clumsy to have the negative
of one momentum physical, the dividend is that some results assume an
elegant form, making it easy easy to permute vertex legs and obtain results
in other channels.

Within the covariant formalism, we associate wavefunction $\phi^*(p)$ with an
outgoing and $\phi(p)$ with an incoming particle of momentum $p$. For spin 0
particless we just interpret $\phi \rightarrow 1$; for spin $\frac{1}{2}$ we
use the traditional Dirac particle wavefunctions $u^{(\lambda)}(p)$ and
$\bar{u}^{(\lambda)}(p)$ interpretation, with normalization
$\sum_\lambda u^{(\lambda)}(p)\bar{u}^{(\lambda)}(p) = \gamma\cdot p + m$;
with spin one we understand $\phi$ to be the standard polarization vector,
$\varepsilon^{(\lambda)}_\mu(p)$, having $\sum_\lambda
\varepsilon^{(\lambda)*}_\mu(p)\varepsilon^{(\lambda)}_\nu(p)
= -g_{\mu\nu}+p_\mu p_\nu/m^2$, dropping the last term for photons
(when amplitudes obey gauge constraints). With excited baryon states 
we regard $\phi$ as the Rarita-Schwinger wavefunction; so for spin 3/2 we 
take $\phi$ to be $u^{(\lambda)}_\mu(p)$, with normalization
$$\sum_\lambda u^{(\lambda)}_\mu(p)\bar{u}^{(\lambda)}_\nu(p) =
\left( -g_{\mu\nu} + \frac{p_\mu p_\nu}{m^2}\right)(\gamma\cdot p + m) -
\frac{1}{3} \left(\gamma_\mu + \frac{p_\mu}{m}\right)(\gamma\cdot p - m)
\left(\gamma_\nu + \frac{p_\nu}{m}\right). $$
Last, but not least, the spinorial wavefunction for an outgoing antifermion
is written as $v^{(\lambda)}(p)$, with normalization 
$\sum_\lambda v^{(\lambda)}(p)\bar{v}^{(\lambda)}(p) = \gamma\cdot p - m$,
and an analogous expression for the spin 3/2 case exists. It is formally true
that the antifermion wavefunction $v^{(\lambda)}(p) = (-1)^{j+\lambda} 
u^{(\lambda)}(-p)= iC\tilde{\bar{u}}^{(\lambda)}(p)$, can be deduced by
appropriate continuation from the fermion wavefunction---a result which is
comparable to that applying to polarization vectors, viz.
$\varepsilon^{(\lambda)}(-p)=(-1)^{j-\lambda}\varepsilon^{(-\lambda)*}(p)$.
This leads to simple substitution rules, as is well-known from field theory,
and we shall take advantage of that fact in the text.

When constructing covariant $\cal M$-functions, to be contracted over external
wavefunctions, we must first count how many independent couplings are needed.
One may deduce their number by the traditional $\l-s$ method or by performing
a helicity amplitude analysis (see below). A more crossing-symmetric
procedure is to count how many ways an {\em integer} angular momentum can
be obtained by combining spins ${\bf j}_1, {\bf j}_2, {\bf j}_3$, since this
does not commit one to any particular channel. The result of such an
analysis is as follows: if one can construct a Euclidean triangle with sides
$j_1,j_2,j_3$ out of the spins, the number of independent couplings is 
$N=j_1(1-j_1)+j_2(1-j_2)+j_3(1-j_3)+2(j_1j_2+j_2j_3+j_3j_1)+1$; Otherwise reorder
the spins in order of decreasing magnitude, $j_a\geq j_b\geq j_c$, whereupon
$N=(2j_b+1)(2j_c+1)$, with $j_a\geq j_b+j_c.$  (Because we are mainly
considering weak processes, parity conservation is a secondary issue.)
Having settled on the number of vertex amplitudes that are needed, we have to
determine appropriate tensors (and matrices for spinors) which they multiply.
This is where a certain amount of freedom exists; for the equations of motion
obeyed by the external wavefunctions allow us to transmogrify a tensor/matrix
form into a combination of other forms. A good example of this freedom is to 
be found in the interaction between a vector meson and two spin 1/2 baryons.
Sandwiched between wavefunctions $\bar{u}(p')$ and $u(p)\varepsilon^\mu (p'-p)$,
the vectors arising in the ${\cal M}_\mu$-function are connected in four ways:
\begin{eqnarray*}
 (p+p')_\mu &=& (m+m')\gamma_\mu + i\sigma_{\mu\nu}(p-p')^\nu \quad 
 ({\rm Gordon~relation}) \\
 (m'-m)\gamma_\mu &=& i\sigma_{\mu\nu}(p+p')^\nu, \quad
 (\rightarrow 0 \quad {\rm for~equal~masses}) \\
 (p-p')^2\gamma_\mu &=& i(m+m')\sigma_{\mu\nu}(p-p')^\nu +
 2\epsilon_{\mu\alpha\beta\rho}p^\alpha p'^\beta\gamma^\rho\gamma_5\\
 (p+p')^2\gamma_\mu&=&(m+m')(p+p')_\mu + i(m'-m)\sigma_{\mu\nu}(p+p')^\nu -
 2\epsilon_{\mu\alpha\beta\rho}p^\alpha p'^\beta\gamma^\rho\gamma_5\\
\end{eqnarray*}
(There are similar relations for axial mesons.) We have``experimented'' with 
various tensors/matrices and derived what we believe is the most elegant 
irreducible set in the main body of the text. This can be confirmed by deriving 
the helicity amplitudes in terms of the covariant ones and verifying their
simplicity. An added advantage of the covariant formalism is that
kinematic threshold factors arise naturally once the contractions over
wavefunctions are carried out in any given channel. Gauge invariance
conditions on the tensors of course lead us to discard certain couplings---in
the helicity formalism it is equivalent to ignoring projections onto
certain helicity states (longitudinal for photons, right-handed for
neutrinos).

The connection with the helicity state formalism is obtained as follows. 
In the centre of mass frame of the decaying particle, with the relative 
3-momentum of the outgoing particles inclined at angle $\theta$ with respect 
to the spin quantization axis---which we are always free to choose---the 
$T$-matrix amplitude is expressed as
$$\langle p_2 \lambda_2; p_3, \lambda_3| T | -p_1,\lambda_1\rangle \equiv
(2\pi)^4\delta^4(p_1+p_2+p_3)
d^{j_1}_{\lambda_1 \lambda_2-\lambda_3}(\theta) M_{\lambda_2,\lambda_3},$$
wherein $M_{\lambda_2,\lambda_3}$ is the reduced matrix element of the
decay amplitude and the rotation matrix $d$ obeys the symmetry properties,
$$d^j_{\lambda\lambda'}(\theta) = 
  (-1)^{\lambda'-\lambda}d^j_{\lambda'\lambda}(\theta) =
  d^j_{-\lambda'\,-\lambda}(\theta) =
  d^j_{\lambda'\lambda}(-\theta).$$
Tables of these $d$ can be found in ref \cite{PD}. It is worth remarking that
all helicity values of the decay products may not be possible in view of 
the fact that $|\lambda_2-\lambda_3|$ cannot exceed $j_1$. In the main
body of the paper we obtain the connection between the covariantly-defined
couplings and the reduced helicity amplitudes; these should not be
excessively complicated if the covariant decompositions are properly chosen 
and indeed serves as a useful guide on how they should be carried out.

So far as discrete space-time symmetries are concerned, if parity happens
to be conserved,
$$M_{\lambda_2,\lambda_3} = \eta_1\eta_2\eta_3 (-1)^{j_2+j_3-j_1}
  M_{-\lambda_2,-\lambda_3}.$$
Time reversal invariance gives no useful information for decays, nor does 
charge conjugation invariance, unless the decay products are uncharged and
self-conjugate. However if the final pair (2 \& 3) of particles are 
identical, we must ensure that the spin-statistics relation, 
$M_{\lambda_2,\lambda_3}=(-1)^{j_1-2j_2}M_{\lambda_3,\lambda_2}$ is satisfied.
Finally we stress again that certain helicity values may be absent for 
massless particles.

Observables are obtainable from the $M_{\lambda,\lambda'}$. One of the most
important of these is associated with the sum over all spins,
$\sum_{\lambda_2,\lambda_3} |M_{\lambda_2,\lambda_3}|^2$, since the partial
decay width for  $1\rightarrow 2+ 3$ is found by averaging over the initial 
spins and including a phase-space factor for the final outgoing pair,
$$ \Gamma = \frac{\Delta}{16\pi(2j_1+1) m_1^3} \sum_{\lambda_2,\lambda_3} 
|M_{\lambda_2,\lambda_3}|^2 .$$
Here the symmetrical Kallen triangle function,
$$\Delta = \sqrt{m_1^4+m_2^4+m_3^4-2m_1^2m_2^2-2m_2^2m_3^2-2m_3^2m_1^2}
 = 2m_1q = 2\sqrt{(p_2\cdot p_3)^2-m_2^2m_3^2}, $$
is proportional to the relative momentum $q$ between 2 \& 3 in the centre of
mass frame of 1 (or in other crossed channels, by permutation of labels)
\footnote{In this connection, recall that powers of $\Delta^{2\ell+1}$
which arise from spin summations are connected with the relative orbital
momentum $\ell$ of the final particles in the centre of mass frame of the
decaying particle.}.
Other observables are connected with polarizations and come through density
matrices of the type $\rho_{\lambda'\lambda''} \propto \sum_\lambda 
M_{\lambda,\lambda'}M^*_{\lambda,\lambda''}$ which often require analysis
of the subsequent interactions or decays of one of the outgoing particles.
An alternative way of extracting individual amplitudes is to examine the
full angular distributions of the decay products or, with less precision,
various asymmetry parameters.

\acknowledgements
This research was supported by a University of Tasmania Special Grant.

\end{document}